\documentclass[aps,twocolumn]{revtex4}
\usepackage[dvips]{graphics,graphicx}
\usepackage{amsmath}
\begin{document}

\title{Bistability and chaos-assisted tunneling in the dissipative quantum systems}
\author{Andrey~R.~Kolovsky$^{1,2}$}
\affiliation{$^1$Kirensky Institute of Physics, 660036 Krasnoyarsk, Russia}
\affiliation{$^2$Siberian Federal University, 660041 Krasnoyarsk, Russia}
\date{\today}
\begin{abstract}
We revisit the problem of quantum bi- and multi-stability by considering the dissipative Double Resonance Model. For a large driving frequency, this system has a simpler phase structure than the driven dissipative nonlinear oscillator -- the paradigm model for classical and quantum bistability. This allows us to obtain an analytical estimate for the lifetime of quantum limit cycles. On the other hand, for a small driving frequency, the system is much reacher than the nonlinear oscillator. This allows us to address a novel phenomenon of dissipation- and chaos-assisted tunneling between quantum limits cycles. 
\end{abstract}
\maketitle

\section{Introduction}
\label{sec1}

Bistability is a widespread phenomenon which one meets in a variety of classical systems ranging from mechanical systems and electric circuits to psychological and biological systems. The paradigm mechanical system showing bistability is the driven dissipative nonlinear oscillator which has two stationary solutions (limit cycles) in a certain parameter region \cite{Land76}. It is naively expected that the driven dissipative quantum nonlinear oscillator also should show bistability. It was found, however, that quantum bistability differs from the classical one. Namely, in the quantum case only one of two limit cycles is the stationary solution while the other cycle is a metastable solution \cite{Drum80,Risk87,Voge90,Bort95,Rigo97}.

In this work we analyze the quantum bistability from a more general perspective of the dissipative nonlinear resonance \cite{10}. Indeed, the two limit cycles in the driven dissipative nonlinear oscillator originate from two nonlinear resonances in the undamped case. Thus, one should meet bi- or multi-stability in any dissipative system whose hamiltonian dynamics supports nonlinear resonances. We illustrate this statement by analyzing the Double Resonance Model (DRM) which is one of paradigm models in the field of classical and quantum chaos \cite{Zasl72,1,2}. For a large driving frequency, this system has a very simple phase-space structure which is well suited for studying quantum limit cycles and their lifetimes. On the other hand, for a small driving frequency, DRM shows a transition to the chaotic regime where the remnants of the nonlinear resonances are seen as stability islands embedded in a chaotic sea. In this case, we find the phenomenon of dissipation- and chaos-assisted tunneling between quantum limit cycles. This relates the considered in this work problem to the problem of chaos-assisted tunneling in the hamiltonian systems which attracted much attention in the past two decades \cite{Toms94,Zakr98,Stec01,Lock10}. We show that dissipation drastically increases the rate of chaos-assisted tunneling and makes it more predictable, that may find useful applications in the field of quantum control.

\section{Double Resonance Model} 
\label{sec2}  

The Hamiltonian of classical DRM reads 
\begin{equation}
\label{a1}
H=\frac{GI^2}{2} - V_{+}\cos(\theta-\omega t) - V_{-}\cos(\theta+\omega t)  \;,
\end{equation}
where $\omega$ is the driving frequency, $G$ the nonlinearity, and $V_{+}=V_{-}=V$ is the perturbation strength.  As a physical realization of DRM one may consider a polar molecule in an alternating electric field. Then $G$ is given by the inverse moment of inertia of the molecule and $V$ by the product of the electric field amplitude and the molecule dipole moment. 

\begin{figure}
\includegraphics[width=8.5cm,clip]{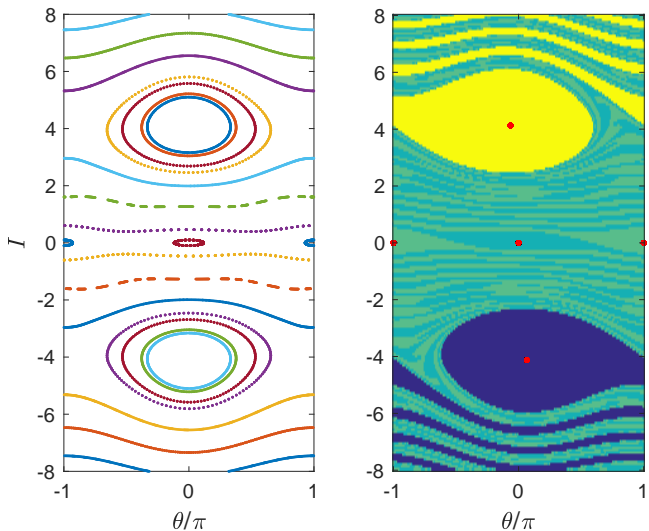}
\caption{Left: Phase portrait of the classical DRM for $\omega=4$, $G=1$, and $V_{+}=V_{-}=1$. Right: The basins of four attractors for $\gamma=0.05$. The red dots mark the positions of the attractors.}
\label{fig1}
\end{figure}

\begin{figure}
\includegraphics[width=8.5cm,clip]{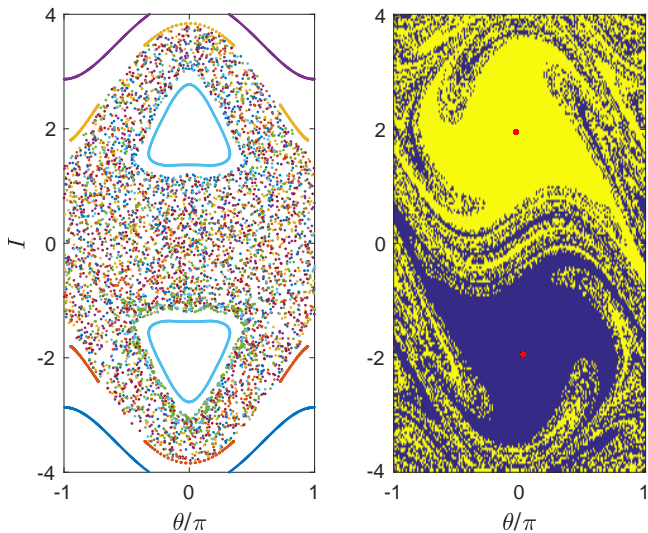}
\caption{The same as in the previous figure yet $\omega=1.6$.}
\label{fig2}
\end{figure}

The system (\ref{a1})  has two primary nonlinear resonances of the width $\delta I=4\sqrt{V/G}$ located at $I_{\pm}=\pm\omega/G$. If the distance between these resonances is much larger than their width, then each of resonances is locally described by the effective Hamiltonian obtained from (\ref{a1}) by setting $V_{+}$ or $V_{-}$ to zero. As an example, the left panel in Fig.~\ref{fig1} shows the stroboscopic map of DRM for $\omega=4$ and the other parameters equal to unity. The two primary nonlinear resonances are clearly seen.   One also finds two secondary nonlinear resonances at $I=0$ and $\theta=0,\pi$ which are due to mutual influence of the primary resonances. If we decrease the driving frequency, this mutual influence becomes stronger that results in the appearance of chaotic separatrix layers which eventually merge into a chaotic sea,  see left panel in Fig.~\ref{fig2}. This chaotic sea first absorbs the secondary resonances at $\omega\approx2$ and then the remnant of two primary resonances at $\omega\approx 0.5$.  

We proceed with the dissipative DRM whose dynamics are governed by the following equations,
\begin{eqnarray}
\label{a2}
\dot{\theta}=\frac{\partial H}{\partial I} \;, \quad
\dot{I}=-\frac{\partial H}{\partial \theta} -\gamma I \;,
\end{eqnarray}
where $\gamma$ is the relaxation constant (the rate of phase-volume contraction). In the case of well-separated resonances dissipation converts two primary resonances into the limit cycles, $I(t)=I_{\pm}$ and $\theta(t)=\theta_0\pm \omega t$, where the relative phase $\theta_0$ is determined by the equation 
\begin{equation}
\label{a3}
V_\pm\sin(\theta_0)=\gamma I_{\pm}  \;,\quad I_{\pm}=\pm \omega/G \;. 
\end{equation}
The secondary resonances transform into the fixed-point attractors where $I(t)=0$ and $\theta(t)=0,\pi$. Relaxation to these attractors can be easily visualized by considering the ensemble of classical particles with initial conditions uniformly distributed over the phase space. This analysis shows that relaxation to the fixed-point attractors goes in two steps -- first, the particles attract to the line  $I=0$  which then shrinks to two points. The right panels in Fig.~\ref{fig1} and Fig.~\ref{fig2} show the basins of the discussed attractors for the relaxation constant $\gamma=0.05$.

\section{Quantum analysis}
\label{sec3}

The quantum counterpart of the Hamiltonian (\ref{a1}) reads \cite{2}
\begin{equation}
\label{b1}
\widehat{H}=\frac{G\hat{I}^2}{2} - V_{+}\cos(\theta-\omega t) - V_{-}\cos(\theta+\omega t)  \;, 
\quad \hat{I}=-i\hbar \frac{{\rm d}}{{\rm d} \theta} \;,
\end{equation}
where $\hbar$ is the effective Planck constant. In the numerical simulations we use the basis 
\begin{equation}
\label{b2}
|n\rangle= (2\pi)^{-1/2} e^{in\theta} \;,\quad  n=0,\pm 1,\ldots
\end{equation}
where the Hamiltonian (\ref{b1}) is given by the three-diagonal matrix. The governing master equation for the system density matrix $\hat{\rho}(t)$ has the form
\begin{equation}
\label{b3}
\frac{d \hat{\rho}}{dt}=-\frac{i}{\hbar}[\widehat{H},\hat{\rho}]  + \widehat{{\cal G}}_{+}(\hat{\rho})+\widehat{{\cal G}}_{-}(\hat{\rho})\;,
\end{equation}
where the Lindblad relaxation operator ${\cal G}_{+}(\hat{\rho})$,
\begin{equation}
\label{b4}
\widehat{{\cal G}}_{+}(\hat{\rho})=-\frac{\gamma}{2\hbar}
(\hat{a}^\dagger\hat{a}\hat{\rho}-2\hat{a}\hat{\rho}\hat{a}^\dagger + \hat{\rho}\hat{a}^\dagger\hat{a}) \;,\quad
\end{equation}
ensures relaxation to the ground state $|0\rangle$ for positive $I$ and the operator  ${\cal G}_{-}(\hat{\rho})$ for negative $I$.

To find the stationary solution of the master equation (\ref{b3}) we rewrite it in the form 
\begin{equation}
\frac{{\rm d}\hat{\rho}}{{\rm d} t}=\widehat{L}(t)\hat{\rho} \;,
\label{c1}
\end{equation}
where $\widehat{L}(t)$ is the linear super-operator. Notice that the super-operator periodically depends on time, $\widehat{L}(t+2\pi/\omega)=\widehat{L}(t)$. Thus, by stationary solution we mean the solution where the density matrix is periodic in time with the same period. Using Eq.~(\ref{c1}) we calculate the Floquet super-operator $\widehat{U}$,
\begin{equation}
\widehat{U}=\widehat{\exp}\left(\int_0^{T} \widehat{L}(t) {\rm d} t \right)  \;, \quad T=\frac{2\pi}{\omega} 
\label{c5}
\end{equation}
(here the hat of the exponent sign denotes the time ordering) and diagonalize it,
\begin{equation}
\widehat{U}\hat{\rho}^{(j)}=\lambda_j  \hat{\rho}^{(j)} \;. 
\label{c6}
\end{equation}
Numerically, this is done by truncating the density matrix  to a finite size $N\times N$ and constructing the column vector of the length $N^2$ by re-ordering the matrix elements $\rho_{n,m}$ in the column-wise manner. Then the super-operator  $\widehat{U}$ is given by the matrix $U$ of the size $N^2 \times N^2$ and the matrices   $\hat{\rho}^{(j)}$ are obtained  by re-ordering the eigenvectors  of this matrix  back to the  $N\times N$ square matrices.

\begin{figure}
\includegraphics[width=8.5cm,clip]{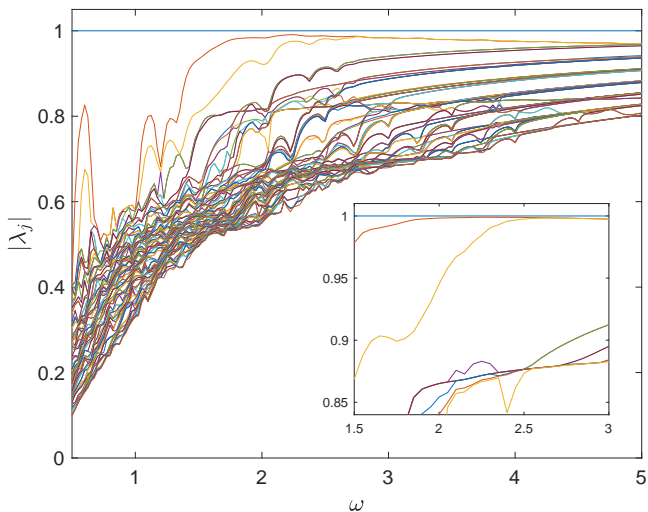}
\caption{Main panel: The spectrum of the Floquet super-operator as the function of the driving frequency $\omega$ for $\hbar=0.5$.  Inset: A fragment of the spectrum for $\hbar=0.25$.}
\label{gig1}
\end{figure}

The main panel in Fig.~\ref{fig1} shows the eigenvalues $\lambda_j$ by the absolute value as the function of the driving frequency $\omega$ for $\hbar=0.5$. Our particular interest in this figure is the stationary state associated with $\lambda_0=1$ and the metastable states associated with the next two levels, which becomes almost degenerate for $\omega>2.5$. Notice that with a decrease of the effective Planck constant $\hbar$ these levels closely approach the level $\lambda_0$ in the certain interval of $\omega$, see inset in Fig.~\ref{gig1}.
In the next paragraph we discuss the stationary and metastable states of the quantum disssipative DRM in more detail.

\begin{figure}
\includegraphics[width=8.5cm,clip]{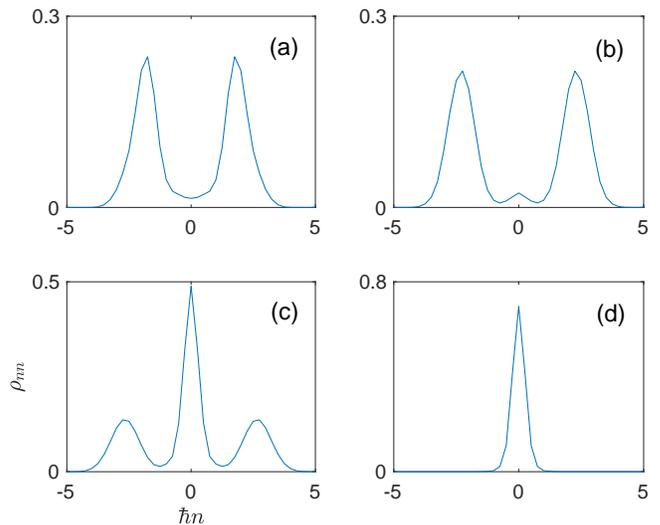}
\caption{Diagonal elements of the stationary density matrix for different values of the driving frequency $\omega=1.6$ (a), $\omega=2.1$ (b), $\omega=2.5$ (c), and $\omega=3.0$. The value of effective Planck constant is $\hbar=0.25$.}
\label{gig2}
\end{figure}

\begin{figure}
\includegraphics[width=8.5cm,clip]{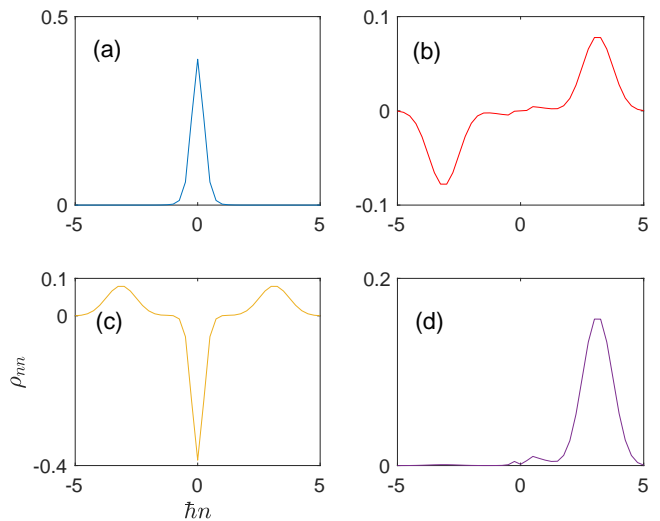}
\caption{Diagonal elements of the density matrices  $\hat{\rho}^{(0)}$, panel (a),   $\hat{\rho}^{(1)}$, panel (b),  and $\hat{\rho}^{(2)}$, panel (c), for $\omega=3$. Panel (d) shows the linear superposition of these matrices which corresponds to the upper limit cycle.}
\label{gig3}
\end{figure}

Fig.~\ref{gig2} shows the diagonal elements of the stationary matrix $\hat{\rho}^{(0)}$ for four different values of the driving frequency. It is seen that the stationary matrix well reproduces the phase-space structure of the classical dissipative DRM. In particular,  one sees two limit cycles in panels (a-c) and the fixed-point attractor at $I=0$ in panels (b-d). There are no limit cycles for $\omega=3.0$; however, they are found in the metastable states $\hat{\rho}^{(1)}$ and $\hat{\rho}^{(2)}$, see Fig.~\ref{gig3}(b,c).  Thus, by using the liner superposition of the first three states one can construct the density matrix which corresponds either to the upper (plus sign) or lower (minus sign) limit cycle, see Fig.~\ref{gig3}(d). The time evolution of this matrix obviously obeys the equation,
\begin{equation}
\hat{\rho}(m T)=\hat{\rho}^{(0)} \pm \lambda_1^m \hat{\rho}^{(1)} +  \lambda_2^m \hat{\rho}^{(2)} \;,
\label{c7}
\end{equation}
where $m$ is the discrete time. Since $\lambda_1\approx \lambda_2$ for $\omega>2.5$,  Eq.~(\ref{c7}) describes the decay of the upper or lower cycle into the fixed-point attractor.  

\section{Decay time of the metastable states}
\label{sec4}

Let us discuss the decay time of the limit cycles originated from primary nonlinear resonances. Assuming the case of well-separated resonances ($\omega>2.5$) we can simplify the problem by setting either $V_{+}$ or $V_{-}$ to zero in the Hamiltonian (\ref{a1}). To be certain, we consider the upper cycle, i.e., $V_{-}=0$. In this case the system has only  two attractors -- the limit cycle at $I=I_{+}$ and an extended simple attractor at $I\approx 0$, which is given by the phase trajectory of the nonlinear resonance with the mean action equal to zero. (Notice that the latter attractor disappears when the phase trajectory becomes the separatrix trajectory.) The relaxation of the system to the discussed limit cycle is locally governed by the equations \cite{10}
\begin{eqnarray}
\label{a4}
\dot{\vartheta}=\frac{\partial H_{eff}}{\partial J} \;,\quad
\dot{J}=-\frac{\partial H_{eff}}{\partial \vartheta} -\gamma J \;,
\end{eqnarray}
where $J=I-I_{+}$, $\vartheta=\theta-\omega t$, and the effective Hamiltonian 
\begin{equation}
\label{a5}
H_{eff}= G\frac{J^2}{2} - V_{+}\cos\vartheta +\gamma I_{+}\vartheta \;.
\end{equation}
The effective Hamiltonian (\ref{a5}) allows us to introduce the local basin of the limit cycle, which we define as the phase volume $S$ encircled by the separatrix trajectory of the Hamiltonian (\ref{a5}).  It follows from Eq.~(\ref{a5}) that the local basin shrinks to zero if $\gamma$ is increased above the critical value which is deduced from Eq.~(\ref{a3}). Indeed, Eq.~(\ref{a3}) has the real solution only if $|\gamma I_{+}/V_{+}| \le 1$. In the opposite limit  of small $\gamma$ the size of the local basin is approximately given by $S=(1/2\pi)\sqrt{V/G}$. 
We mention that the quantum version of Hamiltonian (\ref{a5}) formally coincides with the Hamiltonian of the Wannier-Stark system (a quantum particle in a periodic potential subject to a static force). As known, the Wannier-Stark states are metastable states  \cite{43}.  This fact alone tells us that in the presence of dissipation the quantum nonlinear resonance should have a finite lifetime as well. An estimate for the lifetime of the limit cycle due to the under-barrier tunneling was given in Ref.~\cite{10}. 
\begin{figure}
\includegraphics[width=8.5cm,clip]{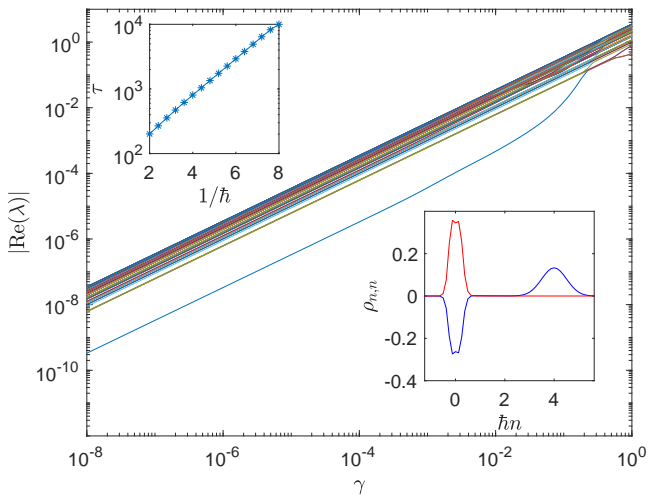}
\caption{Main panel: The first 100 eigenvalues of the operator $\widehat{L}$ as the function of $\gamma$  for $V_{+}=1$ and $V_{-}=0$. The other parameters are $G=1$, $\omega=4$, and  $\hbar=0.25$. Lower inset: Diagonal elements of the matrices $\hat{\rho}^{(0)}$ (red solid line) and  $\hat{\rho}^{(1)}$ (dashed blue line). Upper inset: The lifetime $\tau$ as the function of the inverse Planck constant.}
\label{fig4}
\end{figure}

Since the effective Hamiltonian does not depend on time, the quantum analysis of the problem can be done in terms of the super-operator $\widehat{L}$ (i.e., without constructing the Floquet operator).  The main panel in Fig.~\ref{fig4} shows the real parts of the operator eigenvalues $\epsilon_j$   for $1\le j\le 100$ as the function of the relaxation constant $\gamma$.  It is seen  that for small $\gamma$ the eigenvalue $\epsilon_1$ is well separated from the other eigenvalues whose real parts are approximately equal to $-\gamma$. Thus the system dynamics  for $t\gg 2\pi/\gamma$  is determined by the equation
\begin{equation}
\hat{\rho}(t)=\hat{\rho}^{(0)} + \exp(-t/\tau) \hat{\rho}^{(1)}  \;, \quad \tau=\frac{2\pi}{|{\rm Re}(\lambda_1)|} \;.
\label{c3}
\end{equation}
The lower inset in Fig.~\ref{fig4} shows the diagonal elements of the matrices $\hat{\rho}^{(0)}$ and  $\hat{\rho}^{(1)}$. Thus, Eq.~(\ref{c3}) describes the decay of  the  limit cycle within the characteristic time $\tau$ which, according to the depicted numerical results, is estimated as
\begin{equation}
\tau \sim  \frac{1}{\gamma} \exp\left(A\frac{S}{\hbar}\right)  \;, 
\label{c4}
\end{equation}
where $S$ is the phase volume of the local basin and $A$ a numerical factor.  Notice that in the semiclassical limit $\hbar\rightarrow0$ the lifetime $\tau$ of the discussed quantum limit cycle becomes exponentially large.

The exponential pre-factor in Eq.~(\ref{c4}) is typical for a tunneling process. Yet, there is an alternative interpretation of the finite lifetime due to the intrinsic  quantum noise \cite{Rigo97}. This noise formally appears in the problem if one unravels the master equation for the density matrix into the state diffusion model \cite{Gisi92}. The noise `kicks' the system  out of the basin of the limit cycle attractor into the basin of the fixed-point  attractor that results in a gradual decay of the former.  We find this interpretation also useful to explain the phenomenon of the dissipation- and chaos-assisted tunneling, which we discuss in the next section.

\section{Dissipation and chaos assisted tunneling}
\label{sec5}

Let us discuss the case $\omega< 2$ where the non-dissipative classical DRM  has a large chaotic component with the embedded stability islands, see the left panel in Fig.~\ref{fig2}. Similar to the case of well-separated resonances, the dissipation `transforms' these stability islands (which are remnants of two primary nonlinear resonances) into limit cycles. However, due to unstable hamiltonian dynamics, the basins of these limit cycles acquire a fractal structure where the basins `penetrate' each other in both the upper and lower half-planes of the phase space, see the right panel in Fig.~\ref{fig2}. Thus, the intrinsic noise will presumably  equilibrate  populations of the limit cycles. 

\begin{figure}
\includegraphics[width=8.5cm,clip]{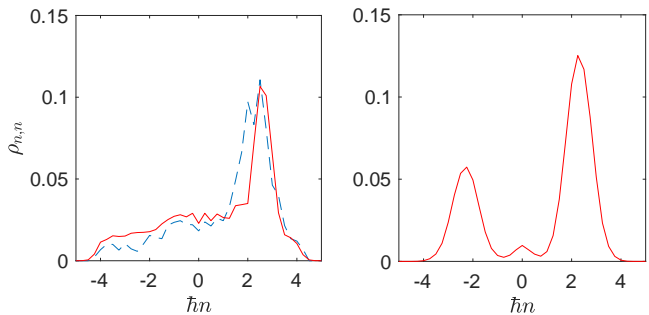}
\caption{Left panel: Diagonal matrix elements of the density matrix at $t=300T$ (solid lines) as compared to the classical distribution function for the action variable (dashed lines). The system parameters are  $V_{-}=V_{+}=1$, $\omega=2.1$,  $\gamma=0$, $\hbar=0.25$. Right panel: Diagonal matrix elements of the density matrix at $t=300T$ for $\gamma=0.05$. In this case, the stationary classical distribution function is given by three $\delta$-peaks of the hight $0.01,0.02,0.97$ located at $I=-\omega,0,\omega$, respectively.}
\label{fig5}
\end{figure}

The above conjecture is fully supported by the straightforward numerical simulation of the system dynamics according to master equation (\ref{b3}). As the initial condition, we choose population of the single level with the quantum number $n_0=I_{+}/\hbar$.  (In the classical case this initial condition corresponds to the ensemble of particles with $I(t=0)=I_{+}$ which are uniformly distributed over the phase $\theta$.) The solid red line in the left panel in Fig.~\ref{fig5} shows the diagonal elements of the quasi-stationary density matrix for $\gamma=0$. This quasi-stationary distribution is well approximated by the classical  distribution for  the ensemble of classical particles, see the blue dashed line. Here the right peak is associated with particles captured into the upper stability island and the background with particles in the chaotic component. Notice the absence of the left peak, which is due to the fact that classical particles cannot escape out or penetrate in a stability island.  These processes, however, are allowed for a quantum particle, where the effect of tunneling is seen in Fig.~\ref{fig5}(a) as higher than in the classical case background  and smaller (narrower) stability island  peak. 

Now we switch on dissipation. For $\gamma\ne 0$ the overwhelming majority of classical particles from the initial ensemble are attracted to the upper limit cycle and stay there forever. In the quantum case, however, we observe probability leakage from the upper to lower cycle, see the red solid line in the right panel in Fig.~\ref{fig5}.  This equilibration process is described by the equation similar to Eq.~(\ref{c3}) where, however,  $\hat{\rho}^{(0)}$ and  $\hat{\rho}^{(1)}$ are now symmetric and antisymmetric matrices with respect to the inversion $n\rightarrow -n$. In the course of time, the `antisymmetric' solution  $\hat{\rho}^{(1)}$ decay  that results in two slightly breathing (with the frequency $\omega$) peaks of the equal heights.   We stress that relaxation to the stationary state is orders of magnitude faster than in the case $\omega>2.5$, where the classical DRM has no chaotic component.

\section{Conclusion}

We analyzed quantum limit cycles in the dissipative DRM. One of the two main results of the work is Eq.~(\ref{c4}) which gives lifetime $\tau$ of the quantum limit cycle as the function of the effective Planck constant.  It should be stressed that  Eq.~(\ref{c4}) is valid only in the case of large driving frequency $\omega$ where the limit cycles are far from each other. In the case of small driving frequency, the decay of limit cycles into the fixed-point attractors at $I=0$ is `substituted' by the dissipation- and chaos-assisted tunneling between limit cycles.  It is found  that dissipation greatly   enhances the tunneling rate as compared to the rate of chaos-assisted tunneling in the non-dissipative DRM.

In the present work we restrict the analysis of the dissipative DRM to the values of the driving frequency $\omega>1.5$, where the relaxation time $\tau$ to the stationary state is a smooth function of $\omega$.  If we go to smaller $\omega$, the classical DRM  shows a sequence of bifurcations where the number of attractors and their types (including chaotic attractors \cite{36}) abruptly change. As a consequence, the relaxation time $\tau=\tau(\omega)$ shows erratic fluctuations and it would be interesting to look at the parameter region $\omega<1.5$ more attentively.

The other direction of research is analysis of the quantum dissipative DRM by using the pseudoclassical approaches \cite{Carm99,116} which substitute the master equation for the density matrix by the Fokker-Planck equation for the classical distribution function. These methods allow one to consider much smaller effective Planck's constants and, thus, to study quantum-classical correspondence in more detail.


\end{document}